\documentstyle{article}
\begin{document}
\renewcommand{\baselinestretch} {1.5}
\large

\vskip 1.0in
\begin{center}
{\large{\bf Quantum Monte Carlo Study of the Hubbard Ladder\\}}
\vskip 1.0in
   Yasuko Munehisa\\
\vskip 0.5in
   Faculty of Engineering, Yamanashi University\\
   Kofu, Yamanashi, 400 Japan\\
\vskip 1.0in
{\bf Abstract}
\end{center}

We present quantum Monte Carlo results for the Hubbard model
on a ladder using the re-structuring method which employ eigenstates
of two-site system on a rung to construct a complete set.

{}From technical reasons we concentrate on the case in which the hopping
along the leg is much less than the hopping on the rung.
We observe the ground state of the half-filled system is made of
singlet-like states described in this paper and its first excited state
contains one triplet rung.
When the system is lightly doped states composed of an electron and
a hole appear on some rungs.
We find very few hole-pairing on the rung upon doping
up to the quarter-filled case.

\eject
\noindent {\bf Section 1  \ Introduction}

Lately strongly correlated systems of electrons on ladders with
two or more legs have become quite intriguing from both theoretical
and experimental points of view. Various models are numerically
investigated by means of exact diagonalization \cite{DRS,TTR,PTR}
and renormalization group approaches \cite{NWS,FK,ENHPS}.
Active studies on the quantum spin ladder systems, which correspond
to the half-filled fermionic ladders with infinite Coulomb repulsion,
are also stimulative \cite{White}.
Among their results, especially interesting are spin gaps,
ground states as well as excited states of the systems
and formation of the hole-pair bound state.
Presence of spin gaps is reported for several spin systems
and for fermion systems such as the $t$-$J$-$t'$-$J'$ model
on a single ladder (two coupled chains) studied in Ref.~\cite{DRS},
the $t$-$J$ model on a single ladder \cite{TTR} and
on multi-ladders with even number of coupled
chains \cite{PTR}, and the two-chain Hubbard Model \cite{NWS}.
In Refs.~\cite{DRS,TTR} emergence of the hole-pairing upon doping,
which would indicate the high $T_c$ superconductivity, is also claimed.

Quantum Monte Carlo techniques based on the Suzuki-Trotter formula,
which prove useful for many spin systems \cite{Hatano,TYM1},
are difficult to apply to the fermion systems because of the negative
sign problem except for those on a single chain \cite{HSSB}.
Usually studies on the two-dimensional fermion systems have been done
after integrating the fermionic degrees of freedom
\cite{IH,WSSLGS,BM,ZCG}.
Yet it would be desirable to simulate such electron systems keeping
fermionic degrees of freedom and vividly observe how electrons and
holes behave on ladders.
Recently dynamical properties of the half-filled Hubbard ladder are
studied by Scalapino {\it et al.} \cite{ENHPS} using grand canonical
quantum Monte Carlo calculations with maximum entropy analytic
continuation.
In this paper we formulate the Hubbard Model on a single ladder,
with or without doping, applying the re-structuring approach
\cite{TYM1}.
Here we employ a complete set composed of sixteen eigenstates of
electrons sitting two ends of a rung, including the double occupancy
(See the Table).
Although our simulations are limited to the case in which the hopping
along each chain is much weaker compared to the hopping between sites
on a rung, we obtain there Monte Carlo results with satisfactory
statistical accuracy on the system's energy as well as what we call
state distribution in this paper, which is defined in section 2,
sorted according to values of the third component of
the total spin. At half-filling we observe that in the ground state
the singlet-like state, the ninth eigenstate in the Table, is sitting
on all rungs of the system and the first excited state has a clear gap
because of the triplet eigenstate which appears on one rung of the system.
Upon doping the lowest lying state we have observed contains not a
hole-pair eigenstate but some of one-electron eigenstates
in addition to the singlet-like eigenstate. Here we see no spin gap.
All our results are compatible with the
corresponding ones in Ref.~\cite{NWS}. For the half-filled ladder our
observations are also in agreement with the $t$-$J$ results in
Ref.~\cite{TTR}.

In the next section we sketch the model and the method used in this
paper. In section 3 the Monte Carlo results are presented. The final
section is devoted to the summary.

\vskip 0.3in
\noindent {\bf Section 2  \ the Model}

The model we study is the Hubbard ladder
with two legs and $N$ rungs. Its Hamiltonian is
\begin{eqnarray}
\nonumber
    {\cal H} & = & -t_c \sum_{i=1}^{N-1}
    \sum_{\sigma = \uparrow , \downarrow} \left [ \left (
     c^{\dag}_{1\sigma}(i) c_{1\sigma}(i+1)
    +c^{\dag}_{2\sigma}(i) c_{2\sigma}(i+1) \right )
    + {\rm H.c.} \right ]  \\
    & & -t_r \sum_{i=1}^{N}
    \sum_{\sigma = \uparrow , \downarrow} \left [
     c^{\dag}_{1\sigma}(i) c_{2\sigma}(i)
    + {\rm H.c.} \right ]
    +U \sum_{j=1}^{2} \sum_{i=1}^{N}
    n_{j\uparrow}(i) n_{j\downarrow} (i)
\label{eq:H}
\end{eqnarray}
where $c^{\dag}_{j\sigma}(i)$ is the creation operator for an electron
of spin $\sigma$ on the $i$-th rung of the $j$-th leg of the ladder
and $n_{j\sigma}(i) = c^{\dag}_{j\sigma}(i) c_{j\sigma}(i)$.
{}From technical reasons we employ an open boundary condition with
a hole at each edge of the legs (See Fig.~1).
The partition function $Z$ is given by
\begin{eqnarray}
     Z={\rm tr} \lbrace e^{-\beta ({\cal H} - \mu{\cal N})} \rbrace ,
\label{eq:Z}
\end{eqnarray}
with the inverse temperature $\beta$, the chemical potential $\mu$
and
\begin{eqnarray}
     {\cal N}= \sum_{j=1}^{2}  \sum_{i=1}^{N}
     \sum_{\sigma =\uparrow , \downarrow} n_{j\sigma}(i)
\label{eq:N}
\end{eqnarray}

Let us describe our re-structuring approach briefly.
In our representation the states of the system are given by
\begin{eqnarray*}
      \mid \alpha \rangle = \mid a_1,a_2,a_3,...,a_N \rangle , \  \  \
      \mid \beta \rangle = \mid b_1,b_2,b_3,...,b_N \rangle ,
\end{eqnarray*}
where $a_m(b_m)$, which is one of the sixteen eigenstates of the
two-site system shown in the Table, denotes the state
on the $m$-th rung along the chain $A(B)$ in Fig.~1.
We expect this choice is proper at least for strong hopping on
rungs.
It should be noted that $a_m$ and $b_m$ may differ by sign
for the same occupants on the two sites of the $m$-th rung
because the {\em left} site on the rung {\em along the chain $A$}
is the {\em right} site {\em along the chain $B$} and vice versa.

Then we divide the above Hamiltonian (\ref{eq:H}) with even $N$
into four parts,
\begin{eqnarray}
\nonumber
    {\cal H}_{A1}  =  \frac{1}{4} {\cal H}_{R}
     -t_c \sum_{k=1}^{N/2}
    \sum_{\sigma =\uparrow , \downarrow} \left [
     c^{\dag}_{1\sigma}(2k-1) c_{1\sigma}(2k)
    + {\rm H.c.} \right ], \\
\nonumber
    {\cal H}_{A2}  = \frac{1}{4} {\cal H}_{R}
     -t_c \sum_{k=1}^{N/2-1}
    \sum_{\sigma =\uparrow , \downarrow} \left [
     c^{\dag}_{2\sigma}(2k) c_{2\sigma}(2k+1)
    + {\rm H.c.} \right ],  \\
\nonumber
    {\cal H}_{B1}  =  \frac{1}{4} {\cal H}_{R}
    -t_c \sum_{k=1}^{N/2}
    \sum_{\sigma =\uparrow , \downarrow} \left [
     c^{\dag}_{2\sigma}(2k-1) c_{2\sigma}(2k)
    + {\rm H.c.} \right ], \\
    {\cal H}_{B2}  =  \frac{1}{4} {\cal H}_{R}
    -t_c \sum_{k=1}^{N/2-1}
    \sum_{\sigma =\uparrow , \downarrow} \left [
     c^{\dag}_{1\sigma}(2k) c_{1\sigma}(2k+1)
    + {\rm H.c.} \right ],
\label{eq:Hd}
\end{eqnarray}
where
\begin{eqnarray*}
    {\cal H}_{R} \equiv -{t_r} \sum_{i=1}^{N}
    \sum_{\sigma =\uparrow , \downarrow} \left [
     c^{\dag}_{1\sigma}(i) c_{2\sigma}(i)
    + {\rm H.c.} \right ]
    +{U} \sum_{j=1}^{2} \sum_{i=1}^{N}
    n_{j\uparrow}(i) n_{j\downarrow} (i) .
\end{eqnarray*}
Note that the Hamiltonian
${\cal H}_{A1} + {\cal H}_{A2} \
({\cal H}_{B1} + {\cal H}_{B2})$ describes the Hubbard
chain $A(B)$ with the hopping parameters $t_c$ and $t_r/2$
and the Coulomb repulsion $U/2$.

Applying the Suzuki-Trotter formula we obtain
\begin{eqnarray}
\nonumber
   Z_n & = & {\rm tr} \lbrace  [
     e^{-\beta ({\cal H}_{A1} - {\mu}' {\cal N} )/n }
     e^{-\beta ({\cal H}_{A2} - {\mu}' {\cal N} )/n }
     e^{-\beta ({\cal H}_{B1} - {\mu}' {\cal N} )/n }
     e^{-\beta ({\cal H}_{B2} - {\mu}' {\cal N} )/n } ]
     ^n \rbrace \\
\nonumber
   & = & \sum_{\langle \alpha_0 \mid} \cdots
         \sum_{\langle \alpha_{2n-1} \mid}
         \sum_{\langle \beta_0 \mid} \cdots
         \sum_{\langle \beta_{2n-1} \mid} \\
\nonumber
   & & \langle \alpha_0 \mid
    e^{-\beta ({\cal H}_{A1} - {\mu}' {\cal N} )/n }
      \mid \alpha_1 \rangle
      \langle \alpha_1 \mid
    e^{-\beta ({\cal H}_{A2} - {\mu}' {\cal N} )/n }
      \mid \alpha_2 \rangle
      \langle \alpha_2 \mid \beta_0 \rangle \\
\nonumber
    & &  \langle \beta_0 \mid
    e^{-\beta ({\cal H}_{B1} - {\mu}' {\cal N} )/n }
      \mid \beta_1 \rangle
      \langle \beta_1 \mid
    e^{-\beta ({\cal H}_{B2} - {\mu}' {\cal N} )/n }
      \mid \beta_2 \rangle
      \langle \beta_2 \mid \alpha_3 \rangle \\
\nonumber
    & & \; \; \; \vdots \\
\nonumber
    & &  \langle \alpha_{2n-2} \mid
    e^{-\beta ({\cal H}_{A1} - {\mu}' {\cal N} )/n }
      \mid \alpha_{2n-1} \rangle
      \langle \alpha_{2n-1} \mid
    e^{-\beta ({\cal H}_{A2} - {\mu}' {\cal N} )/n }
      \mid \alpha_0 \rangle
      \langle \alpha_0 \mid \beta_{2n-2} \rangle \\
    & &  \langle \beta_{2n-2} \mid
    e^{-\beta ({\cal H}_{B1} - {\mu}' {\cal N} )/n }
      \mid \beta_{2n-1} \rangle
      \langle \beta_{2n-1} \mid
    e^{-\beta ({\cal H}_{B2} - {\mu}' {\cal N} )/n }
       \mid \beta_0 \rangle
      \langle \beta_0 \mid \alpha_0 \rangle ,
\label{eq:Zd}
\end{eqnarray}
where $n$ is the Trotter number and ${\mu}' \equiv {\mu} /4$.

For later use let us introduce a hermitian operator ${\cal O}$ defined by
\begin{eqnarray}
{\cal O} = \sum_{\langle \alpha \mid} \mid \alpha \rangle C_{\alpha}
\langle \alpha \mid ,
\label{eq:O}
\end{eqnarray}
where $C_{\alpha}$ denotes a real number. Since
\begin{eqnarray*}
    \langle \alpha_0 \mid {\cal O} \;
    e^{-\beta ({\cal H}_{A1} - {\mu}' {\cal N} )/n }
      \mid \alpha_1 \rangle
& = & \sum_{\langle \alpha \mid} \langle \alpha_0
  \mid \alpha \rangle C_{\alpha}
\langle \alpha \mid e^{-\beta ({\cal H}_{A1} - {\mu}' {\cal N} )/n }
\mid \alpha_1 \rangle  \\
& = &  C_{\alpha_0}
\langle \alpha_0 \mid e^{-\beta ({\cal H}_{A1} - {\mu}' {\cal N} )/n }
      \mid \alpha_1 \rangle  ,
\end{eqnarray*}
we obtain
\begin{eqnarray}
\nonumber
\lefteqn{ {\rm tr} \lbrace {\cal O} \;
e^{-\beta ({\cal H} - \mu{\cal N})} \rbrace } \\
\nonumber
   & = & \lim_{n \rightarrow \infty}
     {\rm tr} \lbrace  {\cal O} \; [
     e^{-\beta ({\cal H}_{A1} - {\mu}' {\cal N} )/n }
     e^{-\beta ({\cal H}_{A2} - {\mu}' {\cal N} )/n }
     e^{-\beta ({\cal H}_{B1} - {\mu}' {\cal N} )/n }
     e^{-\beta ({\cal H}_{B2} - {\mu}' {\cal N} )/n } ]
     ^n \rbrace \\
\nonumber
  &  = & \lim_{n \rightarrow \infty}
      \sum_{\langle \alpha_0 \mid} \cdots
      \sum_{\langle \alpha_{2n-1} \mid}
      \sum_{\langle \beta_0 \mid} \cdots
      \sum_{\langle \beta_{2n-1} \mid}
      C_{\alpha_0} \times   \\
\nonumber
    & & \langle \alpha_0 \mid
    e^{-\beta ({\cal H}_{A1} - {\mu}' {\cal N} )/n }
      \mid \alpha_1 \rangle
      \langle \alpha_1 \mid
    e^{-\beta ({\cal H}_{A2} - {\mu}' {\cal N} )/n }
      \mid \alpha_2 \rangle
      \langle \alpha_2 \mid \beta_0 \rangle \\
\nonumber
    & &  \langle \beta_0 \mid
    e^{-\beta ({\cal H}_{B1} - {\mu}' {\cal N} )/n }
      \mid \beta_1 \rangle
      \langle \beta_1 \mid
    e^{-\beta ({\cal H}_{B2} - {\mu}' {\cal N} )/n }
      \mid \beta_2 \rangle
      \langle \beta_2 \mid \alpha_3 \rangle \\
\nonumber
     & & \; \; \; \vdots   \\
\nonumber
    & &  \langle \alpha_{2n-2} \mid
    e^{-\beta ({\cal H}_{A1} - {\mu}' {\cal N} )/n }
      \mid \alpha_{2n-1} \rangle
      \langle \alpha_{2n-1} \mid
    e^{-\beta ({\cal H}_{A2} - {\mu}' {\cal N} )/n }
      \mid \alpha_0 \rangle
      \langle \alpha_0 \mid \beta_{2n-2} \rangle \\
    & &  \langle \beta_{2n-2} \mid
    e^{-\beta ({\cal H}_{B1} - {\mu}' {\cal N} )/n }
      \mid \beta_{2n-1} \rangle
      \langle \beta_{2n-1} \mid
    e^{-\beta ({\cal H}_{B2} - {\mu}' {\cal N} )/n }
       \mid \beta_0 \rangle
      \langle \beta_0 \mid \alpha_0 \rangle .
\label{eq:OZd}
\end{eqnarray}

With the partition function $Z_n$ in (\ref{eq:Zd}) we simulate
the system on the $N \times 4n$ checkerboard with the number of
rungs $N$ up to 32 and the Trotter number $n=16$ and 24.
We obtain one new configuration by all possible local updates and
global updates in both the spacial
and the Trotter directions using techniques in Ref.~\cite{TYM2}.
We do not fix the number of electrons on a ladder so that the
system would come to the thermal equilibrium swiftly.

Since one is not free from the negative sign problem with the
formulations stated above, our study is limited to the parameter
regions where the cancellation is not serious. To check it
we measure the $r$ ratio,
\begin{eqnarray}
   r =\frac {Z_{+}-Z_{-}} {Z_{+}+Z_{-}},
\label{eq:r}
\end{eqnarray}
where $Z_{+}(Z_{-})$ is the number of configurations with positive
(negative) weight.
We refer to the part that survived the cancellation
as the whole configurations.
A physical quantity $\langle A \rangle $ is calculated by
\begin{eqnarray}
   \langle A \rangle = \frac{A_{+}-A_{-}} {Z_{+}-Z_{-}},
\label{eq:A}
\end{eqnarray}
with contributions from positively (negatively) signed configurations
$A_{+}$ ($A_{-}$).

The physical quantities we measure are the hole doping $\delta$ and
the system's energy $\langle E \rangle$. We obtain $\delta$ by
\begin{eqnarray}
\delta  =  1 - \frac{\langle N_e \rangle}{2N},
\label{eq:delta}
\end{eqnarray}
$N_e$ being the number of electrons on the ladder.
As for the system's energy, we measure the quantity $G \equiv E -
\mu N_e $ using the standard technique in the quantum Monte Carlo
method
and obtain $\langle E \rangle$ by
\begin{eqnarray}
\langle E \rangle  = \langle G + \mu  N_e \rangle .
\label{eq:E}
\end{eqnarray}

In order to shed some light on the system's wave function
we also measure the frequency distribution for every eigenstate
in the Table,
which is the ratio of the number of rungs on a ladder occupied
by that eigenstate to the total number of rungs $N$ and will be
called state distribution.
It should be noted this quantity is physical too,
because the operator to represent the number of each eigenstate
on a ladder is expressed by (\ref{eq:O}).
For the hole-pair eigenstate, for example,
$C_\alpha$ in (\ref{eq:O}) is zero if none of $a_i$'s
in the state $\mid \alpha \rangle$ are equal to $\mid 00 \rangle$,
is one if $\mid \alpha \rangle$ contains one and only one
$\mid 00 \rangle$ state among its $a_i$'s, and so on.

The third component of the total spin of the system, $S_z$, is
another measurable quantity in this formalism. We will see $S_z$
is very helpful to study system's energy spectra.

\vskip 0.3in
\noindent {\bf Section 3  \  Results}

Now let us show our Monte Carlo results.
Typically first ten thousand configurations are discarded for the
thermalization and
next a few ten thousand configurations are used for the measurement.
Throughout the rest of the paper we take $t_r$=1 and, unless stated
otherwise, $U$=6.

We concentrate on the case $t_c \ll 1$, where
the negative sign problem is not so serious.
In this case the ground state at half-filling is expected
to be made of singlet rungs.
We simulate the system with $t_c = 0.1$, namely $t_r / t_c =10$.
In Fig.~2 we plot the state distribution on the $N=16$ ladder
as a function of $\beta$, with
values of $\mu$ adjusted to ensure $\delta$ small enough
to realize the half-filled system. Values of the $r$ ratio
are between $0.58$ and $1$ for these $\beta$ and $\mu$.
We see the mixed state ($\lambda _1$), the ninth state in the
Table, dominates as $\beta$ increases
to become almost $100 \%$ of the whole at $\beta = 15$.
We also observe $S_z =0$ for $99.6 \%$ of the whole
configurations at this value of $\beta$.
These results indicate we can treat the state at $\beta = 15$ as
the system's ground state.
The ratio of the singlet component in the ground state is $2 u_1^2$
($u_1$ being defined in the Table Caption), which is about $ 0.92 $ for
$U=6$ and goes to $1$ as $U$ goes to infinity.
We refer to this state as the singlet-like state.

In order to study the excitation spectrum at half-filling,
we examine the value of $S_z$ at $\beta = 13$ and $\delta \sim 0$,
where we find that $98.85 \%$ of the whole configurations has $S_z=0$
and $0.59$ $(0.56) \%$ has $S_z=1$ $(-1)$.
The state distribution with non-zero $S_z$ shows that
$6.235$ $(6.244) \%$ of the $S_z=1$ $(-1)$ part
is occupied by the sixth (the eleventh) eigenstate, namely by the
triplet(1) (the triplet($-$1)) state, the rest being occupied by
the singlet-like state. These results strongly indicate that about
$1.73 \%$ of the whole configurations at $\beta=13$
(including the triplet(0) state which should be about $0.58 / 98.85$
of the $S_z =0$ part) is in an excited mode in which one of the
$N(=16)$ singlet-like rungs in the ground state turns to the triplet
one.
By measuring system's energies of $S_z=0$, $1$ and $-1$ configurations
separately we estimate the energy of this excited state, $E_1$, from
energies of $S_z = \pm 1$ configurations and the ground state energy,
$E_0$, from energy of $S_z =0$ configurations and $E_1$.
The energy gap $E_1 - E_0$ thus obtained is $0.608 \pm 0.002$, which
is quite close to the difference of eigenvalues $-2\mu -(-\lambda_1)
\simeq 0.606$.

Similar results are obtained on the ladders with different $N$.
Figure~3 plots the energy gap as a function of $1/N$.
The results for $t_c=0.1$ up to $N=32$ indicate that
the energy gap of this excitation mode, namely the spin gap,
remains finite in the thermodynamic limit. Results for $t_c=0.2 $
up to $N=16$ and for $t_c=0.3$ and $N=8$ plotted together also
support the presence of the gap.
Measurement of the gap for larger ladders or larger $t_c$ was
unsuccessful because of the decreasing value of $r$.

Let us see what happens to the system upon doping, which is done by
reducing the value of the chemical potential.
More cancellation due to the negative sign problem is observed
at intermediate values of $\delta$, yet it is not difficult to obtain
statistically meaningful results except for the region near $\delta
\sim 0.25$.
Fig.~4~ shows state distribution at
$\beta=13$ for various values of $\delta$.
Here we observe, from half-filling
($\delta = 0$) to nearly quarter-filling ($\delta = 0.5$),
the singlet-like state diminishes as $\delta$ grows
and the ratio of the one-electron states increases instead.
As is plotted together, only the second and the fourth eigenstate,
which are even in the site-exchange, appear as the one-electron states.
Note that for $\delta < 0.5$ we see very few hole-pairs on rungs.
For $\delta > 0.5$ the distribution
changes drastically because the number of holes exceeds
the number of electrons, but we pursue this region no longer.

Since we do not fix the number of electrons in the updating processes,
configurations with different numbers of electrons can emerge
in the simulation even when the chemical potential is adjusted
to result in integer values of $\langle N_e \rangle$.
In simulations at half-filling this causes no problem because
we observe that for large values of $\beta$ very few configurations
have $N_e$ different from $2N$.
For the doped system, however, we find the contamination is serious
at any available $\beta$.
In search for the ground state of the system with $2(N-N_0)$ electrons
$(N_0=1,2, \cdots )$, we therefore have to discard configurations with
$N_e \neq 2(N-N_0)$ in the measurements of the state distribution and
the energy spectra at the cost of statistical accuracy.
The results thus obtained at $\beta=9$
for the $N=16$ ladder with $N_0=1$ ($i.e.$ for $ \delta = 0.0625)$
show us that the ratio of the second, the fourth and the ninth
eigenstates in the Table (namely the even one-electron state with
up spin, with down spin and the singlet-like state) is about
$1/16:1/16:7/8$ for $S_z = 0$ whereas it is $1/8:0:7/8$
$(0:1/8:7/8)$ for $S_z=1$ $(-1)$ and contributions from other
states are negligible.
Consistently with these results, we do not see any energy gap here;
the energies are $\langle E \rangle = -10.69 \pm 0.01$,
$-10.67 \pm 0.02$ and $-10.63 \pm 0.02$
for $S_z=0$, $1$ and $-1$, respectively.
It should be noted that these values of $\langle E \rangle$ are smaller
than $-10.48$, the value with $t_c=0$ directly calculated
from the eigenvalues of the rungs.
We also simulate the $N=32$ ladder with $N_0 = 2$ at $\beta =9$ and
obtain statistically meaningful results on $S_z=0, -1$ configurations
which suggest that the system has two (one) even one-electron rungs
with up spin, two (three) even one-electron rungs with down spin and
28 singlet-like rungs for $S_z=0$ ($-1$).
The energy on this ladder is $\langle E \rangle = -21.30 \pm 0.05$
$(-21.29 \pm 0.07)$ for $S_z = 0$ $(-1)$, which indicates that the
finite-size effect is small.

\vskip 0.3in
\noindent {\bf Section 4  \  Summary}

In this paper we carry out Monte Carlo simulations of the Hubbard model
on a ladder by the re-structuring method we proposed in Ref.\cite{TYM1}.
Having constructed the complete set in the Suzuki-Trotter formula
from the eigenstates for two electrons on a rung,
our method works satisfactorily when the hopping along the chain is
much weaker compared to the hopping within rungs.
At half-filling we obtain numerical results which indicate that
the ground state is of singlet-like eigenstate only, the first excited
state contains one triplet rung besides singlet-like ones and the spin
gap persists in the thermodynamic limit.
When the system is doped we observe that on some rungs the eigenstates
which contain one electron and one hole appear instead of
the singlet-like ones seen in the half-filled case.

We did not observe any hole-pair formation in the parameter
region we could study. Whether the doped system
realizes hole-pairing or not on ladders with larger
hopping parameters along the chain is an open question so far.
Although it seems difficult to obtain data for such ladders
straightforwardly, detailed study on
the system's wave function which is possible in this formulation
would be useful to understand properties of the system.

\vskip 0.3in
\noindent {\bf Acknowledgement}

The author is deeply grateful to T. Munehisa for fruitful discussions
and helpful comments.

\eject

\eject
\noindent {\bf Table Caption}

Eigenstates, eigenvalues for the partial Hamiltonian of
two-site system on one rung,
\begin{eqnarray*}
 h - \mu  n_e \equiv -t_r \sum_{\sigma = \uparrow , \downarrow}
(c^{\dag}_{1\sigma} c_{2\sigma} + H.c.)
    +U \sum_{j=1}^{2} n_{j\uparrow} n_{j\downarrow}
    - \mu  \sum_{\sigma = \uparrow , \downarrow}
     \sum_{j=1}^{2} n_{j\sigma},
\end{eqnarray*}
with $t_r$ and $U$ in the Hamiltonian (\ref{eq:H}) and the chemical
potential $\mu$ in (\ref{eq:Z}),
and their names used in this study.

$\mid S_l,S_r \rangle$ in the Table
represents the state on a rung whose left (right) site is occupied by
$S_l(S_r)$, where $S_l(S_r)$ is either $0$ (hole),
$\uparrow$ (an up-spin electron),
$\downarrow$ (a down-spin electron)
or $\updownarrow$ (an electron pair).

Parameters $u_1$ and $u_2$ and eigenvalues $\lambda_1$ and $\lambda_2$
in the ninth and the tenth eigenstates are defined as
\begin{eqnarray*}
u_1 \equiv \frac{1}{2} \sqrt{1+\frac{U}{\sqrt{U^2+16t_r^2}}}, \\
u_2 \equiv \frac{1}{2} \sqrt{1-\frac{U}{\sqrt{U^2+16t_r^2}}}, \\
\lambda _1 \equiv \frac{1}{2} ( 4 \mu - U + \sqrt{U^2 + 16t_r ^2} ), \\
\lambda _2 \equiv \frac{1}{2} ( 4 \mu - U - \sqrt{U^2 + 16t_r ^2} ).
\end{eqnarray*}

\vskip 1.0in
\noindent {\bf Figure Captions}

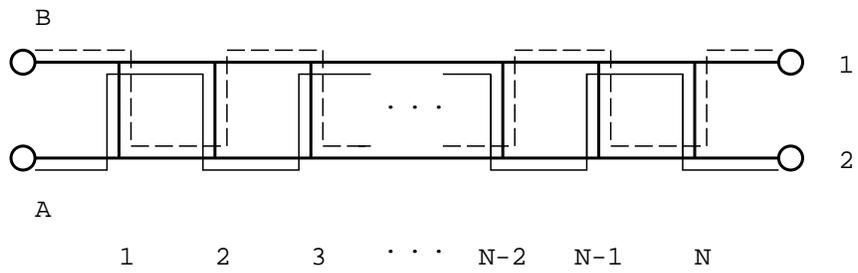
\noindent {\bf Figure 1} \\
A schema to show the ladder we study. The open circle on
each edge of the ladder denotes a hole. The thin solid (dashed)
line is to indicate the chain $A$ $(B)$ described in section 2.

\eject
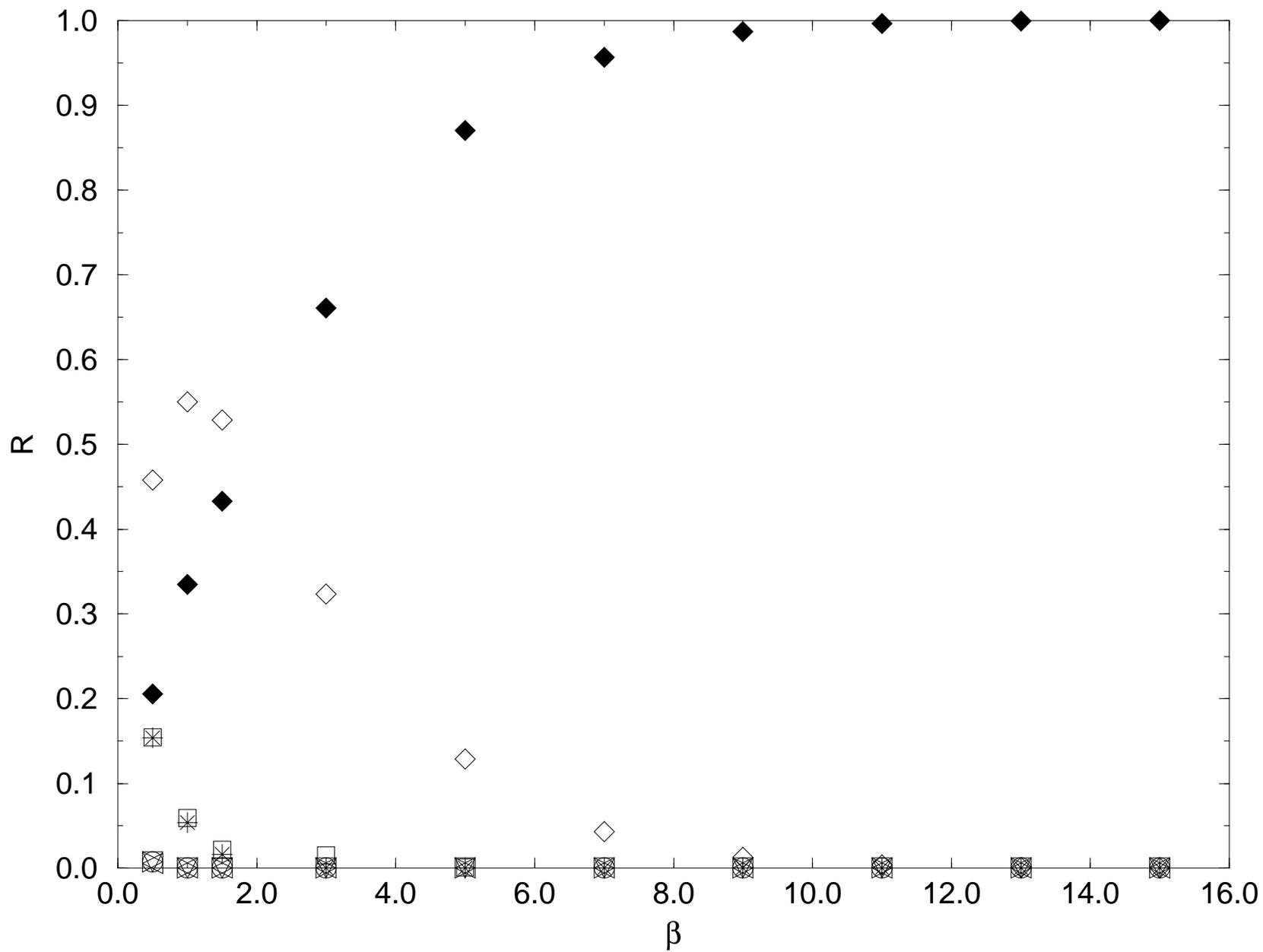
\noindent {\bf Figure 2} \\
The state distribution $R$ for the $t_r=1$, $t_c = 0.1$, $U=6$
and $N=16$ Hubbard ladder as a function of $\beta$ with
$\mid \delta \mid < 7 \times 10^{-3}$. The Trotter number $n$ is 16.
The open circles, right triangles, left triangles and down triangles
denote the ratios for the first, eighth, tenth  and
sixteenth eigenstate in the Table, respectively.
The filled diamonds denote the ninth eigenstate which is
referred as singlet-like state in the paper.
The open squares (asterisks) denote the sum of the
contributions from the second to the fifth (from the twelfth to
the fifteenth) eigenstates, which is the contribution of all
one-electron (three-electron) states.
The open diamonds denote the contribution of all triplet
states, namely the sixth, the seventh and the eleventh eigenstates.
Statistical errors are within symbols.

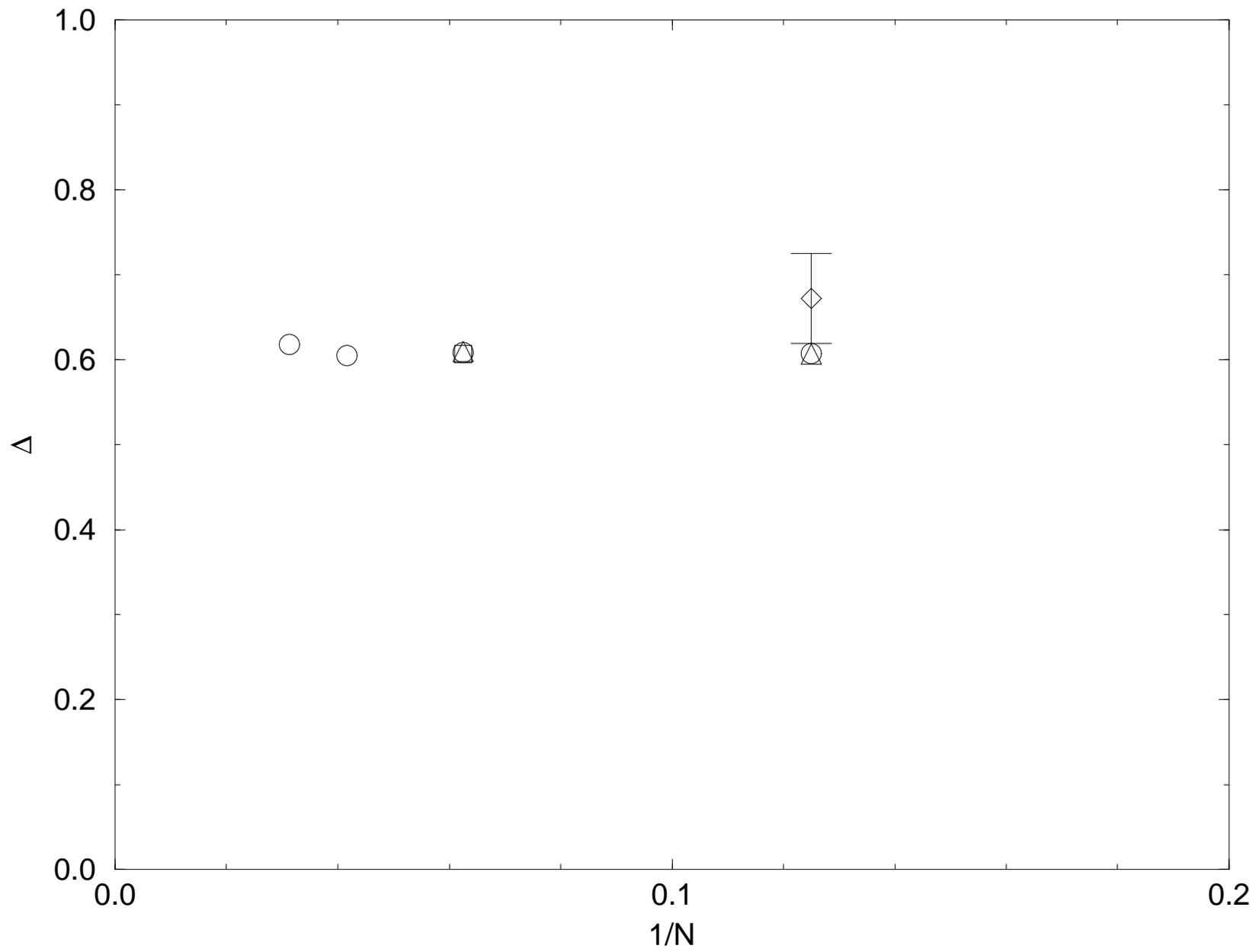
\noindent {\bf Figure 3} \\
The spin gap $\Delta \equiv E_1 - E_0$
at half-filling with $t_r=1$ and $U$=6
as a function of $1/N$, $N$ being the number of rungs of the ladder.
The system's energy defined by (\ref{eq:E}) is measured for the
ground state which
is made of $N$ singlet-like rungs ($E_0$) and for the excited state
where $(N-1)$ rungs are singlet-like and one rung is occupied by the
triplet(1) (or triplet($-1$)) eigenstate ($E_1$).
The open circles denote results for $t_c=0.1$, $\beta =13$
and $n=16$, the open square for $t_c=0.1$, $\beta =13$
and $n=24$, the open triangles for $t_c=0.2$, $\beta=9$
and $n=16$, the open diamond for $t_c=0.3$, $\beta=7$ and $n=16$.
Statistical errors are within symbols except for the $t_c=0.3$ datum.

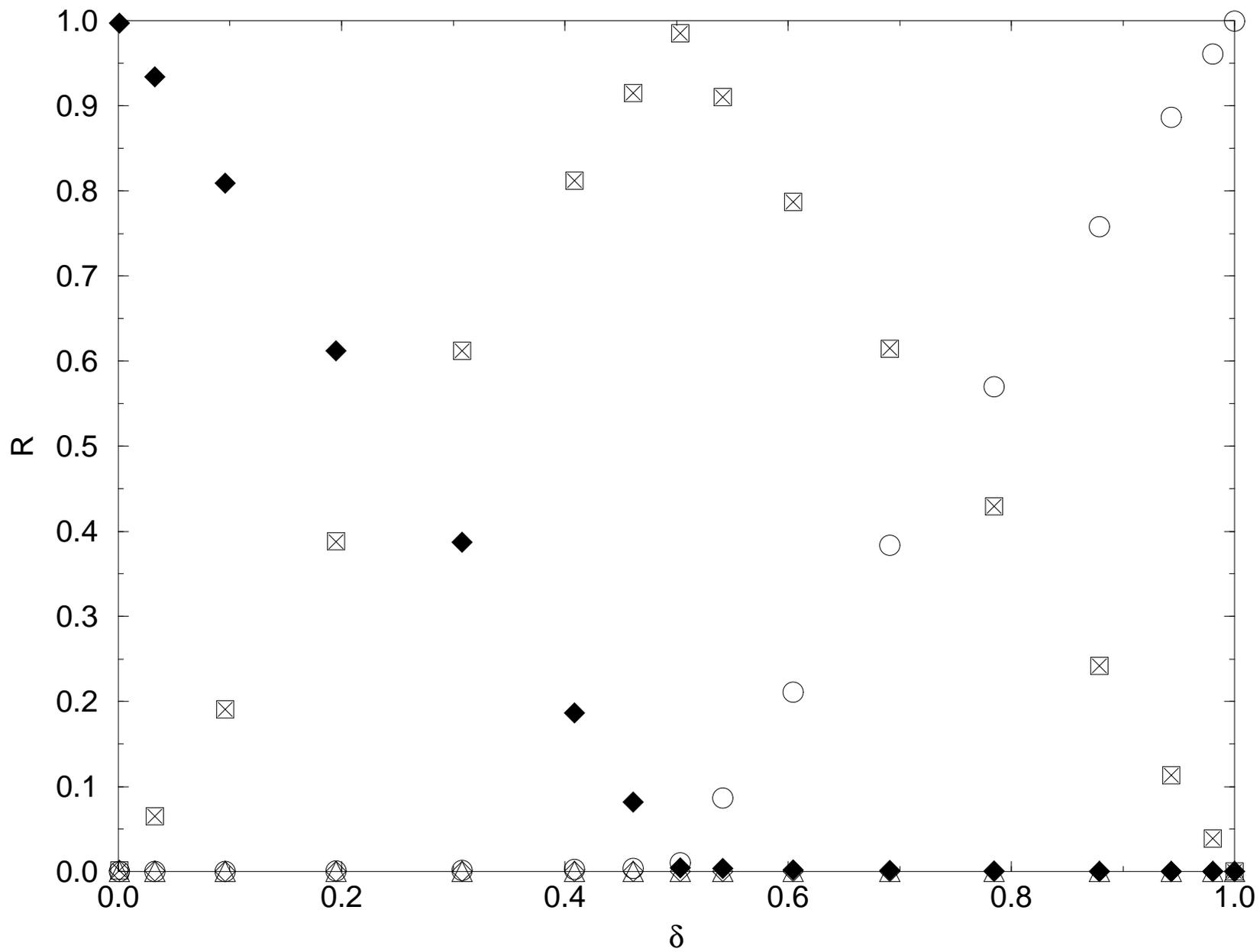
\noindent {\bf Figure 4} \\
The state distribution $R$ for the $t_c = 0.1$, $U=6$ and $N=16$
Hubbard ladder as a function of $\delta$ with $\beta =13$.
The Trotter number $n$ is 16.
The open circles, filled diamonds, open squares and open diamonds
denote contributions of the hole-pair state, the singlet-like state,
all one-electron states and all the triplet states, respectively.
Contribution from the even one-electron states, namely from
the second and the fourth eigenstates, is also plotted by the crosses.
The open triangles are the contribution of the rest.
Statistical errors are within symbols.

\eject
\noindent {\bf Table}

\vskip 0.3in
\begin{tabular}{|c|c|c|c|}  \hline
  $ No. $ &  eigenstate
   & eigenvalue
  &  Name  \\ \hline
  $1$ & $\mid 00 \rangle $
   & 0
  & hole-pair \\ \hline
  $2$
  & $\frac{1}{\sqrt 2}(\mid \uparrow 0 \rangle + \mid 0 \uparrow \rangle ) $
   & $-( \mu  + t_r)$
  & one electron (up,even) \\ \hline
  $3$
  & $\frac{1}{\sqrt 2}(\mid \uparrow 0 \rangle - \mid 0 \uparrow \rangle ) $
   & $-( \mu   - t_r) $
  & one electron (up,odd) \\ \hline
  $4$ & $\frac{1}{\sqrt 2}
  (\mid \downarrow 0 \rangle + \mid 0 \downarrow \rangle ) $
   & $-( \mu  + t_r) $
  & one electron (down,even) \\ \hline
  $5$ & $\frac{1}{\sqrt 2}
  (\mid \downarrow 0 \rangle - \mid 0 \downarrow \rangle ) $
   & $ -(\mu  - t_r )$
  & one electron (down,odd) \\ \hline
  $6$ & $\mid \uparrow \uparrow \rangle $
   & $ -2\mu  $
  & triplet (1) \\ \hline
  $7$ & $\frac{1}{\sqrt 2}
  (\mid \uparrow \downarrow \rangle + \mid \downarrow \uparrow \rangle ) $
   & $ -2\mu  $
  & triplet (0) \\ \hline
  $8$ & $\frac{1}{\sqrt 2}
  (\mid \updownarrow 0 \rangle - \mid 0 \updownarrow \rangle ) $
   & $ -(2\mu  - U)$
  & an electron-pair and a hole  \\ \hline
  $9$ & $u_1
  (\mid \uparrow \downarrow \rangle - \mid \downarrow \uparrow \rangle )
  + u_2
  (\mid \updownarrow 0 \rangle + \mid 0 \updownarrow \rangle ) $
   & $ -\lambda _1$
  & mixed state ($\lambda _1$) \\ \hline
  $10$ & $u_2
  (\mid \uparrow \downarrow \rangle - \mid \downarrow \uparrow \rangle )
  - u_1
  (\mid \updownarrow 0 \rangle + \mid 0 \updownarrow \rangle ) $
   & $ -\lambda _2$
  & mixed state ($\lambda _2$) \\ \hline
  $11$ & $\mid \downarrow \downarrow \rangle $
   & $ -2\mu  $
  & triplet ($-1$) \\ \hline
  $12$ & $\frac{1}{\sqrt 2}
  (\mid \updownarrow \uparrow \rangle + \mid \uparrow \updownarrow \rangle ) $
   & $ -(3\mu  - U -t_r)$
  & three electrons (up,even) \\ \hline
  $13$ & $\frac{1}{\sqrt 2}
  (\mid \updownarrow \uparrow \rangle - \mid \uparrow \updownarrow \rangle ) $
   & $ -(3\mu   - U +t_r)$
  & three electrons (up,odd) \\ \hline
  $14$ & $\frac{1}{\sqrt 2}
  (\mid \updownarrow \downarrow \rangle +
   \mid \downarrow \updownarrow \rangle ) $
   & $-( 3\mu   - U -t_r)$
  & three electrons (down,even) \\ \hline
  $15$ & $\frac{1}{\sqrt 2}
  (\mid \updownarrow \downarrow \rangle -
   \mid \downarrow \updownarrow \rangle ) $
   & $-( 3\mu - U +t_r)$
  & three electrons (down,odd) \\ \hline
  $16$ & $\mid \updownarrow \updownarrow \rangle $
   & $ -(4\mu  - 2U) $
  & electron-pairs \\ \hline
\end{tabular}

\end{document}